\documentstyle[prl,aps,twocolumn,floats,psfig]{revtex}  
\begin{document}

\title{Temperature Dependence of Upper Critical Field as an Indicator
of boson Effects in Superconductivity in Nd$_{2-x}$Ce$_x$CuO$_{4-y}$  }  

\author{V. F. Gantmakher \footnote{e-mail: gantm@issp.ac.ru}, G. A.
Emel'chenko, I. G. Naumenko, and G. E. Tsydynzhapov }

\address{Institute of Solid State Physics RAS, 142432 Chernogolovka, Russia}

\maketitle

\begin{abstract}

The temperature dependence of upper critical field $B_{c2}$ was determined from
the shift of resistive transition $\Delta T(B)$ in nearly optimally doped
Nd$_{2-x}$Ce$_x$CuO$_{4-y}$ single crystals. Within the experimental accuracy,
the weak-field data are described by power function $B_{c2}\propto(\Delta
T)^{3/2}$.  This result is compared with the data on heat capacity and
analyzed in the context of possible manifestations of boson effects in
superconductivity. The $T$ dependence of $B_{c2}$ persists down to the lowest
temperatures, but the numerical values of $B_{c2}$ below 1 K are different for
different samples.


\end{abstract}

There are grounds to believe that high-temperature superconductivity (HTSC) is
not described by the BCS theory. One of them consists in the relationship
between the density $n$ of Cooper pairs and the coherence length $\xi$ (the
pair size). In HTSC cuprates, superconductivity is due to the carriers in the
CuO$_2$ plane.  Like in all 2D systems, the density of states $g_F$ at the
Fermi level in the CuO$_2$ plane does not depend on the carrier concentration
in the normal state and, according to measurements, is equal to
$g_F=2.5\cdot10^{-4}$K$^{-1}$ per one structural unit of CuO$_2$ (this value is
nearly the same for all cuprate families, see, e.g., [1], Ch. 13). Assuming
that the superconducting gap $\Delta$ is of the order of transition temperature
$T_c$, one estimates the mean distance $r=n^{-1/2}\approx(g_F\Delta)^{-1/2}$
between the pairs in the CuO$_2$ plane at 25 \AA\ for $T_c\approx$\,100\,K and
75 \AA\ for $T_c\approx$\,10\,K. These $r$ values should be compared with the
typical coherence length $\xi\approx20$\,\AA\ in the $ab$ plane [1], so that
$r\gtrsim\xi$ in the HTSC materials. Inasmuch as the BCS theory introduces
Cooper pairs to describe the Fermi-liquid ground state as a whole, its validity
for the description of HTSC is not obvious. This causes interest in the models
of Bose superconductivity for which $r\gg\xi$ and which are based on
Bose-Einstein condensation (BEC) in a system of charged bosons [2-4]. The
experimental evidences for the boson effects in HTSC are presently intensively
accumulated.

One such evidence can be expected to obtain from the measurements of the
temperature dependence of magnetic field $B_{c2}$ destroying superconductivity.
In the BCS theory, the $B_{c2}(T/T_c)/B_{c2}(0)$ function is linear in the
vicinity of $T/T_c = 1$; it monotonically increases to saturation near the zero
temperature and almost coincides with the limiting value even at $T/T_c = 0.2$
[5]. However, in most cases, the HTSC materials behave in a different manner
and demonstrate positive second derivative $\partial^2 B_{c2}/\partial T^2$
 over the entire temperature range.

The $B_{c2}(T)$  measurements are mainly based on an analysis of the resistive
 transition. Two types of behavior are known for the resistive transition of
HTSC materials in a magnetic field. For one of them, the transition is sizably
broadened in a magnetic field, so that it is hard and even practically
impossible to gain from it any information about the $B_{c2}(T)$ dependence.
The other transition is shifted in a magnetic field to lower temperatures and
either remains undistorted, as in usual superconductors, or undergoes an
insignificant distortion.  This usually occurs for those members of HTSC
families in which $T_c\le20$\,K. The transition shift in these materials is
naturally explained by the field-induced destruction of superconductivity.
Irrespective of the mechanism of dissipative processes in the superconducting
state, the spectrum rearrangement and the appearance of superconducting pairing
should necessarily affect the $R(T)$ resistance. With this preposition, one can
readily construct the $B_{c2}(T)$ function.

Almost in all HTSC cuprates such as the Tl-based [6] and Bi- based [7] families
and the LaSrCuO [8] and Nd(Sm)CeCuO [9-11] families, as well as in the Zn-doped
[12] or oxygen-deficient [13] YBaCuO, the $B_{c2}(T)$ function derived from the
shift of resistive transition has the positive second derivative over the whole
temperature range $0<T/T_c<1$ and shows a tendency to diverge at small $T/T_c$
values. Most discussion over the $B_{c2}(T)$ curves concentrated precisely on
this divergence and considered it as the most dramatic departure from the BCS
theory. At the same time, the behavior of the $B_{c2}(T)$ function near $T_c$
is also quite informative. Contrary to expectations, almost in all cases where
the field-induced resistive-transition shift in HTSC cuprates pro-ceeds in a
parallel manner, the experimental data indicate that the $\partial
B_{c2}/\partial T$ derivative is zero at the $T_c$ point [6-13].

The $\partial B_{c}/\partial T$ derivative of critical field in the $T_c$ point
is related to the free energy $F$ and heat capacity $C$ in this
point by the well-known Rutgers formula:

\begin{equation}
\frac{1}{4\pi}\left(\frac{\partial B_c}{\partial T}\right)^2_{T_c}=
\frac{\partial^2}{\partial T^2}(F_s-F_n)=\frac{C_s-C_n}{T_c}.
\label{deltaC}
\end{equation}
Inasmuch as the thermodynamic critical field $B_c$ is different from the upper
critical field $B_{c2}$, Eq. (1) can be used only for qualitative estimates.
However, being based on thermodynamics, this equation is very useful.

In usual superconductors, $F_s-F_n\propto(T_c-T)^2$, so that the heat capacity
undergoes a jump and $B_c$ is linear in $(T_c - T)$. In the BEC case,
$F_s-F_n\propto(T_c-T)^3$, so that the heat capacity is a continuous function
in the transition point [14]. It then immediately follows that $\partial
B_{c}/\partial T = 0$ and

\begin{equation}
B_c\propto(T_c-T)^{3/2}.
\label{3/2}
\end{equation}
Of course, one can hardly imagine that the Fermi gas suddenly and completely
transforms into a Bose gas at low temperatures. It was assumed in [4] that
bosons appear in small pockets of the {\bf k}-space near the Fermi level. In
the isotropic model, one can only speak about pairing of sufficiently energetic
fermions, as in the BCS theory. This kind of model has been proposed in [15].
Nevertheless, Eq.  (2) deserves a serious experimental verification.

Such was the motivation of our work consisting in the measurement and analysis
of the field-induced shift of resistive transition in
Nd$_{2-x}$Ce$_x$CuO$_{4-y}$ single crystals. We will discuss separately the
behavior of the $B_{c2}$ field in the vicinity of $T_c$ and at low
temperatures.

{\bf Experiment.}
(NdCe)$_2$ CuO$_4$ single crystals were grown from a mixture of components
taken in the molar ratio Nd$_2$O$_3$:\,CeO$_2$:\,CuO = 1\,:\,0.05\,:\,11 in a
crucible made from yttrium-stabilized zirconium dioxide.  The use of a modified
growth regime markedly reduced the time of interaction between the melt and the
crucible at high temperature. Owing to the accelerated-decelerated crucible
rotation, the melt was intensively stirred so that the homogenization time for
the molten solution did not exceed 1 h at a temperature near 1150 $^\circ$C.
The growth was carried out for several hours upon slow cooling (6 K/h) under
the conditions of morphologically stable crystallization front ($dT/dx \ge
10$~K/cm), after which the crucible was decanted and cooled at a rate of 30-50
K/h to ambient temperature. The crystals were shaped like platelets of
thickness 20-40 mm. Their composition Nd$_{1.82}$Ce$_{0.18}$CuO$_x$ was
determined by local X-ray spectroscopic analysis. The analysis revealed Zn
traces in the crystals at a level of 0.1 wt \%.
Initially the crystals did not show superconducting transition above 4.2 K. The
superconducting transition at $T_c\approx 20\,$K appeared after 15-h annealing
at 900 $^\circ$C in an argon atmosphere.

Measurements were made for two plates approximately $1\times2$~mm in size. The
silver paste contacts were fused in the air at a temperature of
350~$^\circ$C. Four contacts in sample 1 were arranged ~0.5 mm apart in a row
on one side of the plate. The potential contacts in sample 2 were placed on the
opposite side of the plate beneath the current contacts, allowing the measuring
current to be directed both along and transversely the {ab} plane. This did not
affect the results. The resistance was measured by the standard method using a
{\it lock-in} nanovoltmeter at a frequency of 13 Hz. The measuring current was
small enough for the linear regime and the absence of overheating to be
provided down to the lowest temperatures. The magnetic field was directed along
the normal to the plate ($c$ axis). Measurements were performed over the
temperature range from 25 K to 25 mK.
\footnote{The low-temperature measurements in strong magnetic fields were
carried out at the NHMFL (Tallahassee, Fla., USA).}
The onset of zero-field superconducting transition in both samples occurred at
about 20.5\,K.

\begin{figure}
\centerline{\psfig{file=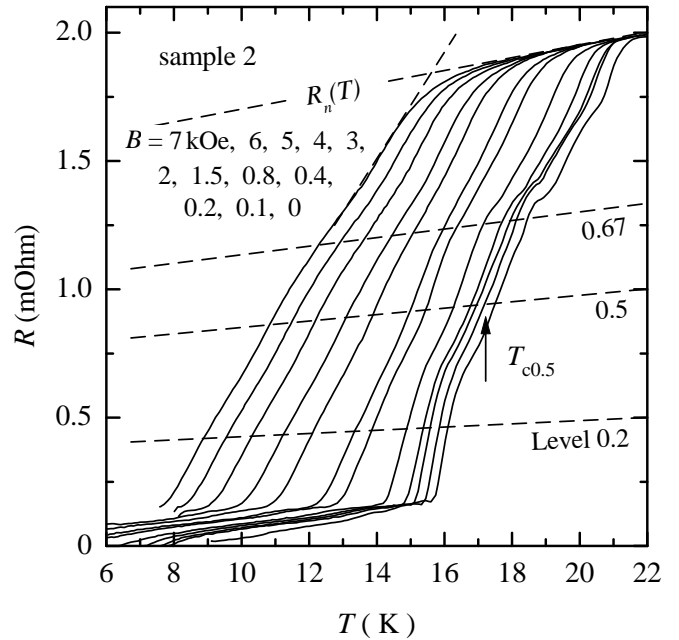,width=\columnwidth,clip=}}
\vspace{3mm}
\caption{
The $R(T)$ curves for sample 2 in magnetic fields
(from right to left) from 0 to 7 kOe. The dashed lines are the
straight line $R_n(T)$ and the straight lines at the levels of 0.67,
0.5, and 0.2 of $R_n(T)$. The method of determining the onset
of transition is demonstrated and the $T_{ci}$ fields from which
the shifts were measured are shown} 
\end{figure}

The measurements gave identical results for both samples. Figure 1 demonstrates
a series of low-field $R(T)$ curves for sample 2. At high temperatures, all
curves show the same asymptotic behavior $R_n(T)$ above the transition, and one
can assume that the $R_n$ function does not depend on $B$ at $T > 10$--12 K.
The zero-field transition shows a certain structure, which, however, is
smoothed out even at 100-200 Oe. The field effect mainly amounts to shifting
the transition to lower temperatures.  The degree in which this shift is
parallel can be checked by comparing the shift of the onset of transition with
the shifts of the $R(T)$ function at different levels: $0.2R_n$, $0.5R_n$, and
$0.67R_n$ (see curves in Fig. 1).  For the parallel shift, all constructions in
Fig. 1 should give the same function $B_{c2}(\Delta T)$, where $\Delta
T=T_{ci}-T$ and $T_{ci}$ is the temperature corresponding to the same level on
the initial curve $R(T, B = 0)$. The log-log plots of the shifts are shown by
different symbols in Fig. 2a for all four levels. The systematic deviations of
the symbols from the straight line

\begin{equation}
B_{c2}=(\Delta T)^\beta,
\label{beta}
\end{equation}
constructed by averaging the results for all points are small for each of the
symbols. This implies that the distortions of the transition shape are small as
compared to its shift. The scatter of points in low fields is mainly caused by
the fine structure of the $R(T, B = 0)$ curve, which serves as a reference in
the determination of the shift $\Delta T$.

\begin{figure}
\centerline{\psfig{file=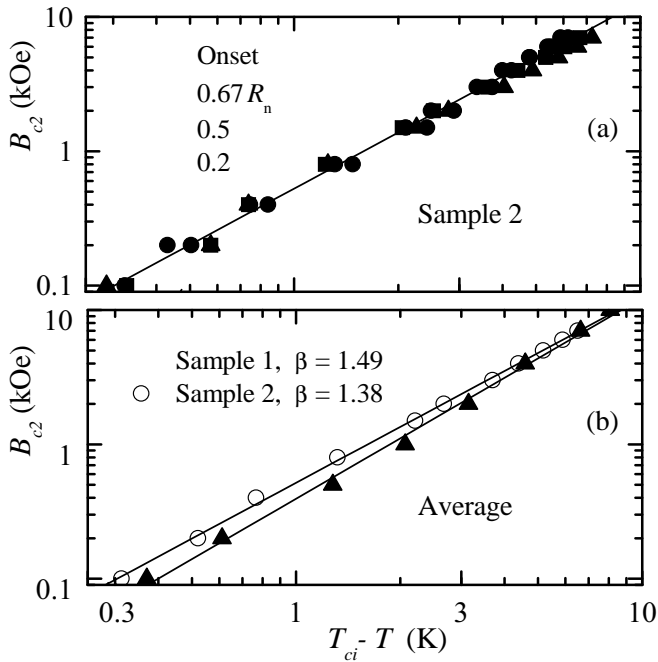,width=\columnwidth,clip=}}
\vspace{3mm}
\caption{
(a) Plots of the field vs. shift at different levels in this
field; (b) the same for the averaged shifts for two samples.}
\end{figure}

The coefficient $\beta$ was determined from the slope of the straight line
passing through the averaged $\Delta T$ shifts (Fig. 2b). The curve processing
for sample 2 (Fig. 1) yields $\beta\approx1.4$, and the processing of analogous
curves for sample 1 yields $\beta\approx1.5$.

The resistances for both crystals decreased in a relatively narrow temperature
range not to zero; one can see in Fig. 1 that, starting at the level of
$\sim0.1$, a slanting tail appears. The same tail for sample 1 starts at a
higher level of $\sim0.2$. In this work, we will analyze only the upper portion
of the transition, assuming that the electron spectrum is rearranged into the
form characteristic of the superconducting state precisely in this region.

\begin{figure}
\centerline{\psfig{file=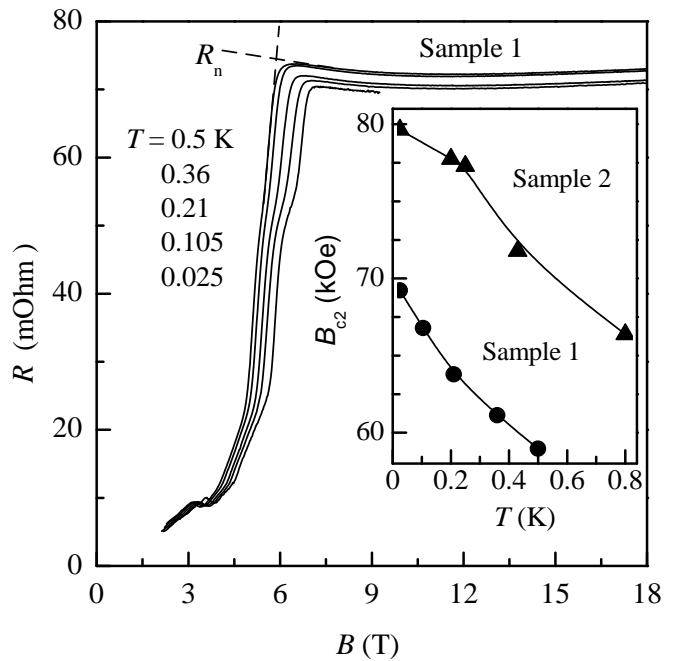,width=\columnwidth,clip=}}
\vspace{3mm}
\caption{
The $R(B)$ curves for sample 1 at temperatures (from
left to right) from 0.5 K to 25 mK. Inset: the field of the
onset of transition at low temperatures for both samples.
}
\end{figure}

Figure 3 shows the $R(B)$ functions for very low temperatures $T/T_{c0}<0.05$.
In this region, the normal resistance depends, though weakly, on a magnetic
field, while the onset of transition is clearly defined and its shift is easily
detected even upon changing temperature below $T/T_{c0}=0.005$. When
considering the $B_{c2}(T)$ functions in this region (see inset in Fig. 3), two
fact are noteworthy. First, $B_{c2}$ does not show tendency to diverge near
zero temperature; although the derivative of $B_{c2}(T)$ is large below 0.5 K,
the function is linear within the experimental accuracy and extrapolated to a
finite value $B_{c2}(0)$ (similar result was obtained previously for thallium
crystals [6]). Second, the critical fields at low temperature are equal to 69
and 80 kOe for samples 1 and 2, respectively, i.e., differ by more than 10\%,
inspite of the fact that the crystals were from the same batch and their $T_c$
values coincide.

\begin{figure}
\centerline{\psfig{file=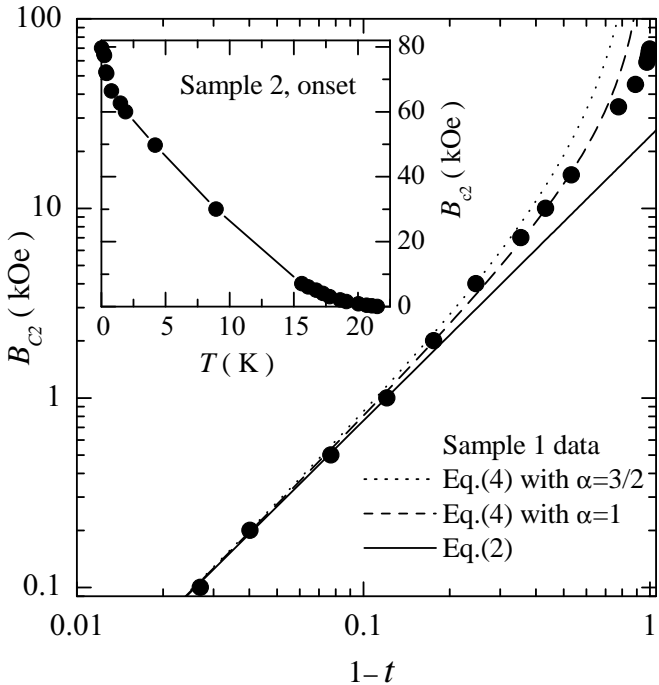,width=\columnwidth,clip=}}
\vspace{3mm}
\caption{
Comparison of the experimental $B_{c2}$ values for sample
1 with Eqs. (2) and (4). Inset: the $B_{c2}(T)$ function for
sample 2 in the whole temperature region. The line is a guide to the eye.}
\end{figure}

The graph of $B_{c2}(T)$ over the entire temperature range is shown in the
inset in Fig. 4; as in other HTSC cuprates, the second derivative $\partial^2
B_{c2}/\partial T^2\ge0$ for all temperatures (cf., e.g., [6, 7]).

{\bf Discussion.} It follows from the preceding section that our data for the
vicinity of $T_c$ are consistent, within the experimental accuracy, with Eq.
(2). It would have been instructive to compare these data with the data on heat
capacity, but, unfortunately, in the works where the heat capacity of
Nd$_{2-x}$Ce$_x$CuO$_{4-y}$ was measured [16] the contribution of critical
fluctuations near $T_c$ was not determined. Nevertheless, it is known that the
measurements of heat capacity of the HTSC materials show strong dissimilarity
from usual superconductors [17] but do not allow the discrimination between the
BCS and BEC models. These problems can be illustrated by comparing the results
of measurements of the resistance and heat capacity of the thallium high-$T_c$
superconductor.  No explicit jump in heat capacity is observed for this
compound even at zero field, although the contribution from the critical
fluctuations is undoubtedly present in the temperature range 16-10 K [18]; this
contribution is reduced by approximately one-half in a field of 0.4 T and
remains virtually in the same temperature range. At the same time, the
resistive measurements by the same experimental group [6] suggest that a field
of 0.4 T shifts the transition by 25\% from 16 to 12 K.

In connection with this contradiction, an interesting remark was made in [19],
where the numerical calculations were carried out for the heat capacity of an
ideal charged Bose gas in a weak magnetic field. It is well-known that the BEC
does not occur in the ideal charged Bose gas in a uniform magnetic field [20]
because the density of states diverges at the lower Landau level of the spectra
of charged bosons. This implies that the transition occurs only at an isolated
point in the $(T, B)$ plane. The magnetic field in this plane is scaled by the
comparison of the cyclotron energy $\hbar eB/mc$ with $T_c$.  Substituting the
free electron charge and mass for $e$ and $m$, respectively, one arrives at the
value of 8 T for the characteristic field at $T_c=16\,$K. On this scale, the
above-mentioned field of 0.4 T is as small as 0.05. As long as the field is
low, the phase trajectory again passes through the vicinity of the transition
point in the $(T, B)$ plane upon changing $T$, but, as the field increases, the
"impact parameter" increases, while the contribution of critical fluctuations
decreases. However, the temperature interval corresponding to the small impact
parameters does not change. In the case that the transition is the BEC in a
weakly nonideal charged Bose gas, this contribution is hidden from view at the
lower temperature where the transition occurs in magnetic field. Then, strange
as it may seem, the resistive measurements provide the more reliable
information on the transition position than the heat capacity measurements do.

According to the results obtained for the immediate vicinity of $T_c$, the
behavior of the $B_{c2}(T)$ function should be compared with the predictions of
the superconductivity models in a nonideal Bose gas. Due to the boson
scattering by impurities or to the boson-boson interaction, the critical field
in a weakly nonideal Bose gas behaves as [21]

\begin{equation}
B_{c2}\propto t^{-\alpha}(1-t^{3/2})^{3/2}, \qquad t=T/T_c,
\label{bipol}
\end{equation}
where, depending on the particular model, the exponent $\alpha$ is equal to 1
or 3/2 [21, 22]. At $t\to1$, function (4) takes the asymptotic form (2). It is
seen in Fig. 4 that the experimental points deviate in the proper direction
from the asymptote and, on the whole, correspond well to Eq. (4). A more
detailed comparison is hardly pertinent, as long as the theories [21, 22] do
not allow for the field-induced pair decay into fermions.

{\bf Conclusions}. The field-induced distortion of the shape of resistive
superconducting transition in the Nd$_{2-x}$Ce$_x$CuO$_{4-y}$ single crystals
is appreciably smaller than the transition shift. This allows the measurement
of the $B_{c2}(T)$ function. As zero-field $T_c$ is approached, the $B_{c2}$
field behaves as a power function $B_{c2}\propto(\Delta T)^\beta$ with
$\beta\approx1.5$ and, correspondingly, with a horizontal tangent $\partial
B_{c2}/\partial T=0$. This should imply the absence of jump in heat capacity at
the zero-field phase transition. Such a behavior is precisely that which is
expected for the heat capacity and critical field in the BEC of a charged Bose
gas. For this reason, one of the possible conclusions that can be drawn from
such a behavior of $B_{c2}(T)$ near $T_c$ is that the description of
superconductivity of the HTSC materials should involve the BEC elements, i.e.,
make allowance for the fact that fermions near the Fermi level tend to form
bosons at temperatures above $T_c$.  The $T$ dependence of $B_{c2}$  persists
down to the lowest temperatures, although, probably, the $B_{c2}$ values in
this region depend on lattice defects.

We are grateful to A.A. Abrikosov, L.P. Gor'kov, and V.P. Mineev for helpful
discussions. The experiments at the NHMFL were performed in the framework of
the program of cooperation between the NHMFL and scientists from the former
USSR. This work was supported by the Russian Foundation for Basic Research
(project no. 99-02-16117), the Russian Foundation for Basic Research---PICS
(project no. 98-02-22037), the State Contract 107-2(00)-P, and the program
"Statistical Physics" of the Ministry of Sciences of the Russian Federation.

\end{document}